\documentclass[11pt]{article} 
\usepackage{makeidx}
\usepackage{graphicx}
\usepackage[hang,small,bf]{caption}
\title{A dynamical model of non regulated markets}  
\author{A. Schaale \\ \\  \small{ \it Postfach 900129, D-12401 Berlin, Germany }}  
\begin{document} 
\begin{titlepage}  
\maketitle  
\begin{abstract}  
The main focus of this work is to understand the dynamics of
non regulated markets. The present model can describe the dynamics of any
market where the pricing is based on supply and demand. It will
be applied here, as an example, for the German stock market presented by the
Deutscher Aktienindex (DAX), which is a measure for the market status. 
The duality of the present model consists of the superposition of the two 
components - the long and the short term behaviour of the market. 
The long term behaviour is characterised by a
stable development which is following a trend for time periods of years or
even decades. This long term growth (or decline) is based on on the development of fundamental market figures. 
The short term behaviour is described as a dynamical evaluation (trading) of
the market by the participants. The trading process is described as an
exchange between supply and demand. In the framework of this model there the
trading is modelled by a system of nonlinear differential equations. The model
also allows to explain the chaotic behaviour of the market as well as periods
of growth or crashes.
\medskip 
\\ PCAS numbers: 01.75.+m, 05.40.+j, 02.50.Le 
\medskip
\\ Contribution to the technical seminar 22/12/98, DESY-IfH Zeuthen
\end{abstract} 

\end{titlepage}  
\setcounter{page}{2}
The traditional approaches of pricing models (indices, stocks, currencies,
gold, etc.) are related to combinations of economic figures like profit or cash-flow  and their
expected development. Indeed, these fundamental
figures are related to the approximate price. However, it is well known
that similar objects (companies, goods, ...) can be priced on
the same market quiet different. One can observe quick changes in the
pricing, which can't be explained by any change of the underlying basic
figures \cite{davidson}.    
\medskip

The present model consists of two basic components: 
\begin{itemize}
\item (Long term trend) Scaling of the price (index) based on the long term
      development of basic figures \cite{mandelbrot1} 
\item (Short term trend) Pricing by the exchange between buyers (optimists),
      sellers (pessimists) and neutral market members
\end{itemize}
 Studying, for example, the DAX $I$ for a time period of one
 decade one will recognise, that the basic trend $I_0(t)$ shows an exponential
 behaviour  with deviations (fig. 1.). This trend can be presented as: 
\begin{equation}  
I_0(t)=\hat{I}_0\,e^{\lambda t}, \;\; \mbox{with} \;\; \hat{I}_0, \lambda=const.  
\label{defi0}  
\end{equation}  
The parameter $\hat{I}_0$ is the starting value: 
$\hat{I}_0(t)\equiv I_0(t=0)$.   
The growth rate $\lambda$  can variate on different markets. This parameter
summarises all basic influences on the market, such as economic freedom,
taxes, social-economic parameters, infrastructure and others. Comparing
different markets one will find, that certain economics are growing (US,
Europe) while others are declining over years (Japan \footnote{A long term
decline of national economies is often caused by massive regulations, reducing the
economic freedom.}).
\medskip

The value for the parameter $\hat{I}_0$ and $\lambda$ can be fitted from the
historical market data using the least square method. The development
$I_0(t)$ symbolises the average growth of the economy which is measured in
various economic figures. The growth in (\ref{defi0}) fulfils the Euler
equation, describing the ``natural'' growth of unlimited systems:
\begin{equation}
I_0'(t)-\lambda I_0(t)=0
\label{euler}
\end{equation} 
where $y'(t)\equiv \frac{d}{dt}y(t)$.
The function (\ref{defi0}) describes a real growth process. 
\medskip

As far as there is no universal pricing model, the individual evaluations by
the market participants differ and the price deviates from the fundamental
average. These different evaluations which are changing in time lead to
some kind of spontaneous oscillations. As far as each market has another scale it is useful
to normalise the market index (price) to make different markets better comparable: 
\begin{equation} 
I(t) \rightarrow i(t)=\frac{I(t)}{I_0(t)}
\label{itrans} 
\end{equation}  
The function (\ref{itrans}) performs a normalisation which will project all
indices of real markets to a unitarian index $i$ with a constant basic trend: 
$i_0(t) \equiv 1$ and $\lambda=0$. This way the development of markets can be
compared in a single scheme.
\medskip

For further discussions it is necessary to define the market structure. A
market is the totality of all market members participating in the
trading process \cite{caldarelli}. The total amount of market members on normalised markets
(\ref{itrans}) is constant. The normalised DAX can be found in (fig. 2.).
\medskip 

As already mentioned above, the subjective evaluations of
the market status differ from each other \cite{davidson}. The market participants can be
separated into three groups: optimists, pessimists and neutral market
participants. Each group has a certain concentration which evolves in time
$c_k(t)$. Based on the normalisation there is:
\begin{equation} 
c_o(t)+c_p(t)+c_n(t)=1,  
\label{marktnorm} 
\end{equation}   
with $c_o(t)$, $c_p(t)$ and $c_n(t)$ as the corresponding concentrations 
\footnote{The concentration is the weighted average of individual market members with a 
  similiar market view, but a different capitalization
$$
c_i(t)=\frac{1}{M(t)}\, \sum_{k=1}^{N_i}m_k(t) 
$$
where $i=o,p,n$ represents the corresponding market views of optimists, pessimists
and neutral market members. $m$ is their individual capital and $M$ the summary
market capitalization. $N_i$ is the total
number of individual market members with the same market view.}  
\medskip 

The dynamics of the market is a result of the development of the $c_k(t)$ and
the index $i(t)$. Each market group has certain features and react on market
changes in a different way:
\begin{itemize} 
\item {\bf Optimists} consider the market to be priced low. They want to buy.
\item {\bf Pessimists} consider the market to be priced high. They want to sell.
\item {\bf Neutral market members} consider the market to be priced
           fair. They are passive.  
\end{itemize} 
The groups have different sizes. Comparing a typical daily trading volume 
with the total market capitalisation one will find that it is orders of
magnitude smaller ($<1\%$). This leads to the following relation between the
concentrations:  
\begin{equation} 
c_o(t), c_p(t) << c_n(t). 
\label{relationen}
\end{equation}  
Using (\ref{marktnorm}) the dimension of the problem reduces from 3 to 2
independent functions $c_p(t)$ and $c_p(t)$. The system dynamics can
be written in the form of a system of differential equations:
\begin{equation} 
c_k'(t)= L_k\big(c_o(t),c_p(t),t\big),\;\; k=o,p 
\label{ct-system2}
\end{equation}  
Now it is necessary to describe in $L$ the structure of the market drivers,
which determine the dynamics of trading:
\medskip

On non regulated markets there the price is determined by supply and demand. The
ratio of the concentration of optimists and pessimists defines the price
level \cite{farmer}. In general the functional relation between the
concentrations of different market members and the index $i$ can be expressed
in the following form:
\begin{equation}
i(t)=f\Bigg(\frac{c_o(t)}{c_p(t)}\Bigg).
\label{marktdruck}
\end{equation}
At present it is not possible to derive the explicit form of $f$ from 
economic principles. The function $f$ expresses the {\it
  subjective} evaluations of market participants. 
Here and in the following there will be made extensive use of
Taylors theorem. Unknown functions will be expanded in Taylor series in order
to parametrise them. As far as the higher order terms of each expansions will be
neglected, it is possible to {\it define} the function $f$ in the following
form:
\begin{equation} 
i(t)=f\Bigg(\frac{c_o(t)}{c_p(t)}\Bigg)\equiv \frac{c_o(t)}{c_p(t)}
\label{ip} 
\end{equation}  
In the equilibrium state there the equation  (\ref{ip}) gives sensible
results:
\begin{equation}
c_p(t)=c_o(t) \leftrightarrow i(t)=i_0(t)\equiv 1
\label{gleichgewicht}
\end{equation}
After defining the basic conceptions there will be studied now the
development of the concentrations $c_o(t)$ and $c_p(t)$. Their changes in
time can be expressed by the following system of equations:
\begin{eqnarray} 
c_o'(t) &=& F_o(\Delta i(t))+\xi_o U(t), \nonumber \\
c_p'(t) &=& F_p(\Delta i(t))-\xi_p U(t), \nonumber \\
\Delta i(t) &=& i(t)-1 
\label{defcpprime} 
\end{eqnarray} 
The system describes the exchange of concentrations as functions of the current
index and as external influences. The functions 
$F_k(\Delta i(t)),\; k=o,p$ describe the subjective evaluations of the market
members as a function of supply (pessimists) and demand (optimists). The
function  $U(t)$ represents an ``external field''. It models effects
that influence the market, but which are {\it not} related to the present
value of the index $i(t)$. Typical external influences could be related to interest
rates, taxes, political events or persons. The constants $\xi_k$ describe the
difference in {\it perception} of external influences by the different market
groups. The external influences lead to periods of continues optimism or
depression, as they are observed on real markets. 
\medskip

In general the functions $F_k$ in (\ref{defcpprime}) are unknown. They will be
expanded in Taylor series around the equilibrium state $i_0$:
\begin{equation}
 F_k(\Delta i(t))= \sum_{n=0}^{\infty} \alpha_{k,n}  \cdot \big[\Delta i(t)\big]^n
\label{taylor} 
\end{equation}  
In the following the will be used the following approach: 
\begin{equation}
 F_k(\Delta i(t))= \alpha_{k,1} \Delta i(t) + O([\Delta i(t)]^2).
\label{fp2} 
\end{equation}
In case without external influences $U(t)=0$ it makes sense to assume that the
system is symmetric concerning optimism and pessimism. Otherwise the system
would follow a systematic trend, which has been already taken into account in
(\ref{marktnorm}). This leads to the relation
\begin{equation} 
 F_p(\Delta i(t))=-F_o(\Delta i(t)) \equiv F(\Delta i(t))\; .
\label{fop2} 
\end{equation}   
On ideal markets the perception of external influences would be symmetric
too. Real markets show deviations from this symmetry $\xi_o\neq
\xi_p$. Performing a redefinition of $U(t) \rightarrow \xi_o U(t)$ one can
substitute the the $\xi$ such as $\xi_o=1$ and  
$\frac{\xi_p}{\xi_o}=1+\varepsilon$, where $\varepsilon$ is an empirical
parameter defining the asymmetry of perception of optimists and pessimists. 
\medskip 

Based on several reasonable assumptions, it has become possible 
to construct a nonlinear system of differential equations that reflects  
the market dynamics:
\begin{eqnarray}
 && c'_o(t)-\alpha \Big[c_o(t) c_p^{-1}(t)-1\Big]-U(t)=0,\nonumber \\ 
 && c'_p(t)+\alpha \Big[c_o(t) c_p^{-1}(t)-1\Big]+(1+\varepsilon)\, U(t)=0,
\label{finalsystem} 
\end{eqnarray} 
with the starting conditions: 
\begin{equation} 
 c_o(0)=c_{o0},\;
 c_p(0)=c_{p0}.
\label{finalsystemstart} 
\end{equation} 
and $\alpha \equiv \alpha_1$.
\medskip

The equations of system (\ref{finalsystem}) describe the principal relation
between the concentrations and the market index, where the exchange between
the concentration levels can be performed in {\it infinite small steps}
(continuum limit). That means that the ideal market would react on infinite small
deviations from the equilibrium 
with infinite small trading reactions (exchange of fractions of stocks). This is
not possible on real markets, which react with the exchange of {\it finite
  sized trading units}. This causes discontinuous changes of the index. Each
new trading process is related to the former trading process which itself has
caused a change of the index. One can realize this discontinuous trading behaviour by
transforming the system of differential equations  (\ref{finalsystem}) into  a
system of logistic equations, where the trading process becomes described as a
{\it sequence of finite exchange transactions} \cite{caldarelli,busshaus}:
\begin{eqnarray}
 c_o^{(n+1)} &=& c_o^{(n)}+ \Delta c_o^{(n)}, \nonumber \\
 c_p^{(n+1)} &=& c_p^{(n)}+ \Delta c_p^{(n)}, \nonumber \\
 \Delta c_o^{(n)}  &=& \alpha \Big[c_o^{(n)}
                       \Big(c_p^{(n)}\Big)^{-1}-1\Big] + U^{(n)},\nonumber \\  
\Delta c_p^{(n)}&=& -\alpha \Big[c_o^{(n)}
                    \Big(c_p^{(n)}\Big)^{-1}-1\Big]-(1+\varepsilon)\, U^{(n)}, \nonumber \\ 
 U^{(n)}&\equiv& U(t_n), \nonumber \\ 
  n&=&0,1,... 
\label{finalsystemlog} 
\end{eqnarray} 
with the starting conditions
\begin{equation} 
 c_o^{(0)}=c_o(0),\;\;
 c_p^{(0)}=c_p(0).
\label{finalsystemstartlog} 
\end{equation} 
Now there will be shown the results of the application of the model to real
markets. At first there will be studied growth periods and crashes, which are
observed regularly on all financial markets. Using historical data of the DAX
one can find, that the growth periods are caused 
by an exponential growing external optimism   
\begin{equation} 
U(t)=U_0\,\Big(e^{\beta(t-t_0)}-1\Big)  
\rightarrow U^{(n)}=U_0\Big(e^{\beta(t_n-t_0)}-1\Big)\; . 
\label{uexp} 
\end{equation}
As one can see in (fig. 3.) that an exponential growing external optimism leads to
an exponential growing index $i(t)$. Starting from a certain deviation the
system starts to generate oscillations and becomes instable. This fact may cause pessimism
(or even panic) in a self reinforcing process \cite{bouchaud}.  After some time
this leads to a collapse of the market \cite{hogg}. Therefore crashes are not only the
result of changes in the external 
influences $U$, but they are caused by the internal instability when the
system is far from the equilibrium \cite{caldarelli,johansen, illinski}. Even
if the external optimism would 
continue growing, the system would start to collapse starting from a critical
deviation (DAX: critical deviation at $\pm 35\%$).  An external potential of the type
(\ref{uexp}) is mathematically 
equivalent to a redefinition of the long term trend $I_0(t)$:
\begin{equation}
I_0(t)=\hat{I_0} e^{\beta (t-t_0)} \rightarrow I^*_0(t)=\hat{I^*_0} e^{\beta^*
  (t-t_0)}, \;\;\; \beta^*>\beta
\label{istar}
\end{equation}
This  ``excited'' state exists usually only a certain time period, until the
system reaches the critical deviation. After the begin of the collapse the
external optimism vanishes and the system returns to the equilibrium
state. This behaviour can be found in the historical data of the DAX and other
markets. Phases of continuous growth over several month are followed by phases
of decline. All of these periods show an exponential behaviour. 
\medskip 

The market system is very sensitive concerning changes in the neutral
component of the market $c_n$. Relatively small external influences on the neutral
component become enhanced by a leverage effect on the index. This effect is caused
by the different orders of magnitude of the concentrations (\ref{relationen}):
\begin{equation}
\Delta i(t) \sim \frac{c_n(t)}{c_{p,o}(t)} \, \Delta U(t), \;\;\;
\frac{c_n(t)}{c_{p,o}(t)}\sim 100...1000 
\label{faktor1}
\end{equation}

Another essential feature of the dynamics of markets is the chaotic behaviour,
for example in the daily changes of the index. The reason for the appearance
of chaos is the feedback of the market to itself. The strength of response on
deviations of the equilibrium is described by the model parameter $\alpha$. In
(fig. 4) there are shown examples of the development of the market system
(\ref{finalsystemlog}) in dependence of $\alpha$.
\medskip

In (fig. 4a) there the response of the market is relatively
small, so that the market compensates after several transactions. If  $\alpha$
reaches a critical value (fig. 4b) the reaction on a deviation $\Delta i$ is
that strong, that it creates a new deviation with the same size but opposite
sign. As the result of this the system starts to oscillate. A further increase
of $\alpha$ causes a permanent overcompensation of the market deviations. The
system becomes chaotic (fig. 4c). The parameter $\alpha$ is proportional to
the volatility of markets. 
\medskip

It is worth to remark, that the market shows a typical feature of non linear
problems - fractal patterns. The basic trend over years or decades has an
exponential behaviour. The different fragments (medium term trends) have an
exponential behaviour as well (but a different growth rate).  
\medskip  

In this work the model was applied on financial markets, but it can be
generalised to all markets which are based on supply and demand. The model
describes the long and the short term dynamics of markets within a single
theoretical framework, using a few empirical parameters. The model can
describe crashes as phase transitions, caused by it's internal instability.
Important features of real markets like chaotic behaviour and a fractal
structure are described by a system of non linear differential 
equations. Using this model it is possible to determine basic parameters,
which can describe the status of the market in both, the short and the long
term trend.  
\medskip

I would like to thank Gerhardt Bohm and Klaus Behrndt for their helpful
support and discussion.


\begin{figure}
\begin{center}
\caption{DAX and long term trend $I_0 \;$  since Jan. 3. 1989}
\includegraphics[angle=0, width=150mm]{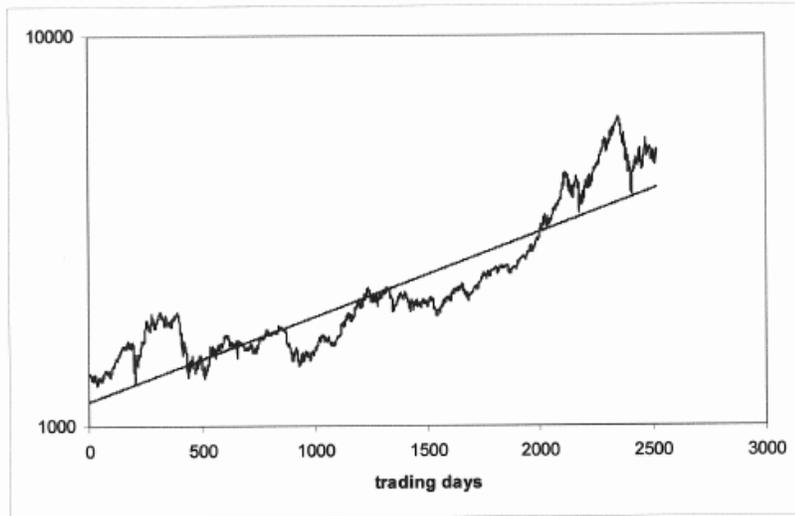} 
\end{center}
\end{figure} 

\begin{figure}
\begin{center}
\caption{DAX and deviation from long term trend $\Delta i \;$  since
  Jan. 3. 1989} 
\includegraphics[angle=0, width=150mm]{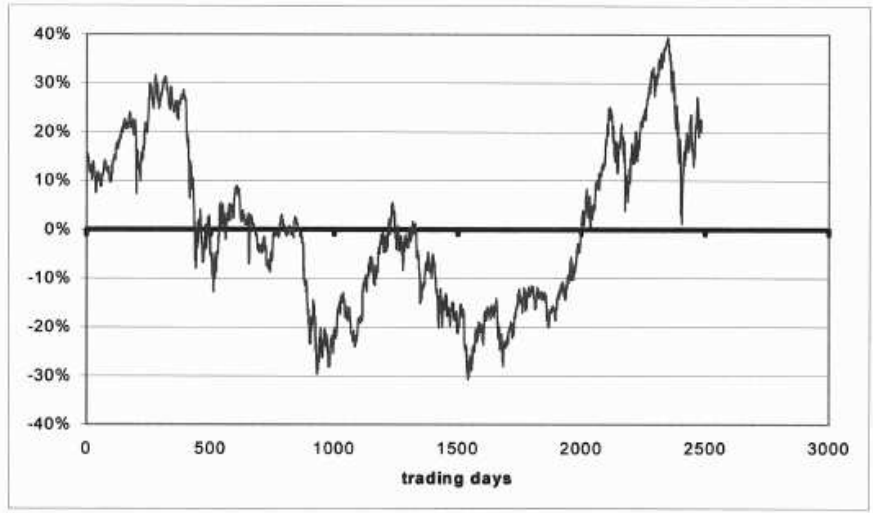} 
\end{center}
\end{figure} 

\begin{figure}
\begin{center}
\caption{Model behavior of $\Delta i$ with external 
   optimism $U_n=U_0
  e^{\beta(t_n-t_0)}$} \vspace{5mm} 
\hspace{2mm}   DAX with medium term trends Jan. 2 1998 - Feb. 4 1999
\includegraphics[angle=0, width=150mm]{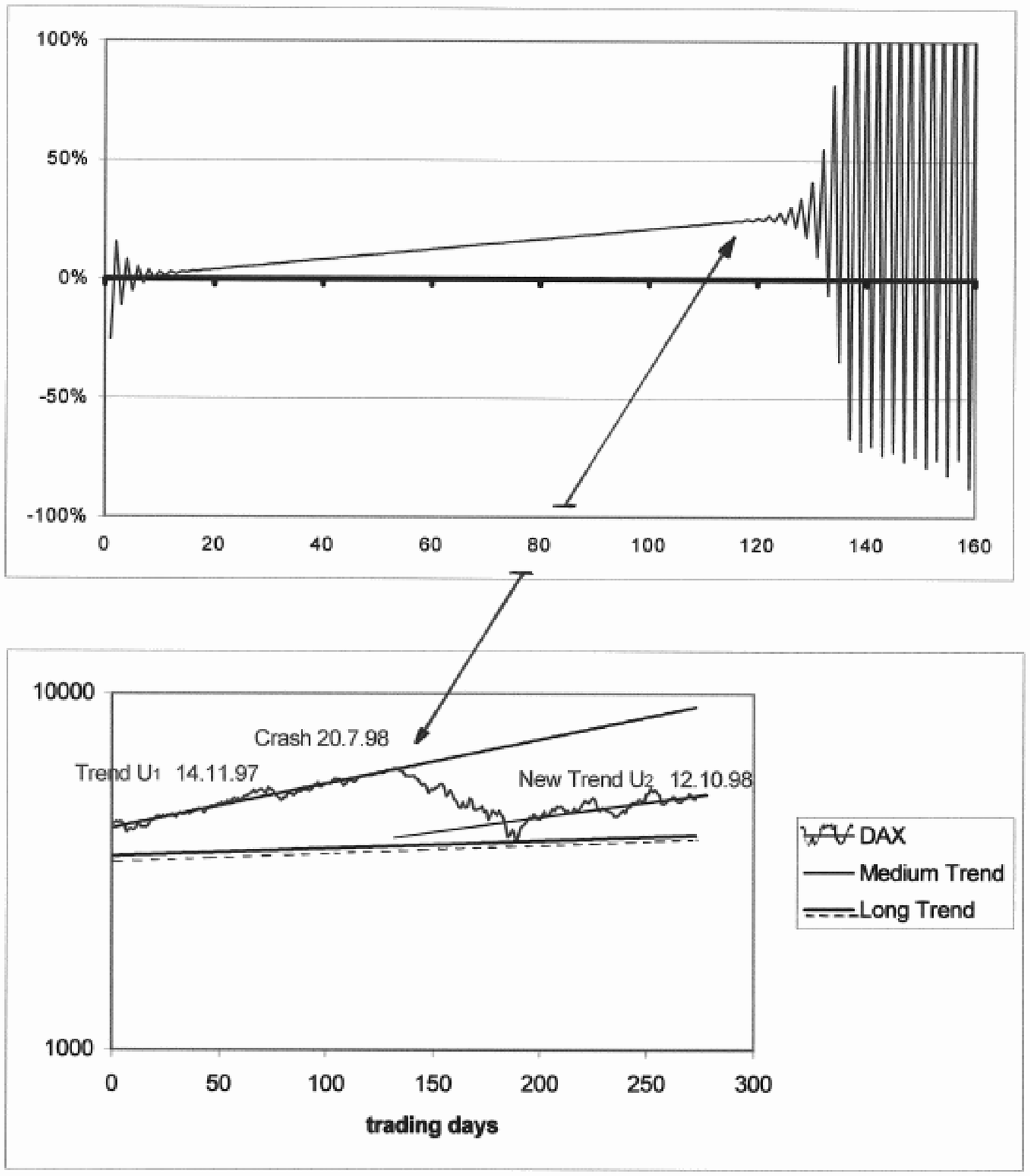} 
\end{center}
\end{figure} 

\begin{figure}
\begin{center}
\caption{Model behavior of $\Delta i_n$ for different parameters $\alpha$}
\includegraphics[angle=0, width=150mm]{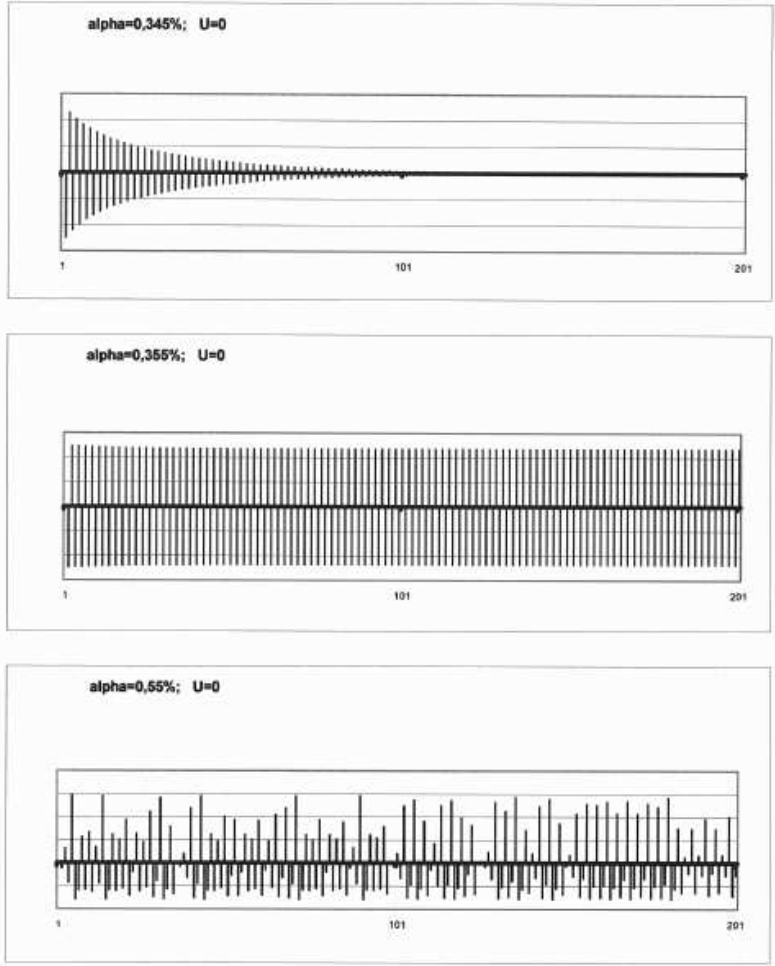} 
\end{center}
\end{figure} 


\end{document}